# Hierarchical Fog-Cloud Computing for IoT Systems: A Computation Offloading Game

Hamed Shah-Mansouri, *Member, IEEE,* and Vincent W.S. Wong, *Fellow, IEEE*

***Abstract*—Fog computing, which provides low–latency computing services at the network edge, is an enabler for the emerging Internet of Things (IoT) systems. In this paper, we study the allocation of fog computing resources to the IoT users in a hierarchical computing paradigm including fog and remote cloud computing services. We formulate a computation offloading game to model the competition between IoT users and allocate the limited processing power of fog nodes efficiently. Each user aims to maximize its own quality of experience (QoE), which reflects its satisfaction of using computing services in terms of the reduction in computation energy and delay. Utilizing a potential game approach, we prove the existence of a pure Nash equilibrium and provide an upper bound for the price of anarchy. Since the time complexity to reach the equilibrium increases exponentially in the number of users, we further propose a near–optimal resource allocation mechanism and prove that in a system with $N$ IoT users, it can achieve an $\epsilon$-Nash equilibrium in $O(N/\epsilon)$ time. Through numerical studies, we evaluate the users' QoE as well as the equilibrium efficiency. Our results reveal that by utilizing the proposed mechanism, more users benefit from computing services in comparison to an existing offloading mechanism. We further show that our proposed mechanism significantly reduces the computation delay and enables low–latency fog computing services for delay–sensitive IoT applications.**

***Index Terms*—Computation offloading, fog computing, Internet of Things, potential games.**

## I. Introduction

### A. Background and Related Work

Fog computing provides cloud services at the edge of the network where data is generated [1]. Fog computing services are promising to alleviate the challenges that Internet of Things (IoT) systems face due to the tremendous growth of IoT devices and applications [2], [3]. Fog computing not only reduces the backbone traffic to be sent to the cloud, but also improves the latency for delay–sensitive IoT applications by reducing the relatively long delay of remote cloud computing. These result in enhanced user–experience [4], [5]. Nonetheless, efficient allocation of fog computing resources is challenging due to the limited processing power of fog nodes and rapid development of computation–intensive applications.

Powerful computing resources are required to support the growing demand of IoT applications with heterogeneous quality of service (QoS) requirements [6]. However, the IoT devices usually suffer from limited processing power. Task offloading to either fog nodes or remote cloud servers can alleviate this issue. IoT devices equipped with multi–radio access technology (multi–RAT) can connect to different fog nodes or remote cloud servers. Each IoT device may select a different fog node or remote cloud server to offload its tasks while guaranteeing the QoS requirement of its applications. To enable fog computing in IoT systems, a resource allocation mechanism is crucial to efficiently allocate the fog computing resources to the IoT devices.

The authors are with the Department of Electrical and Computer Engineering, University of British Columbia, Vancouver, BC, Canada, V6T 1Z4. E-mail: {hshahmansour, vincentw}@ece.ubc.ca.

Task offloading in fog computing has received much attention in recent years due to the growing development of IoT systems. Optimal allocation of offloaded workload in fog-cloud computing is studied in [7]. The objective is to minimize the power consumption of fog nodes and cloud servers when meeting the delay constraints of different IoT applications. An online job dispatching and scheduling mechanism is proposed in [8] that aims to minimize the total weighted response time over all the jobs. The weight is set based on the latency sensitivity of each job. A distributed mechanism for the allocation of fog computing resources is proposed in [9] that aims to minimize the response time of fog nodes under a given power efficiency constraint. This mechanism determines the portion of the computation workload of each fog node that should be offloaded to the remote cloud servers. To provide fog computing services, the formation of fog networks is studied in [10]. An online framework that enables the fog nodes to form a network of computing resources is proposed with the goal of minimizing the maximum delay of all computation tasks generated by all users within the network. A latency-constrained resource allocation mechanism is proposed in [11] when a fog network provides computing services and is able to cache the popular computation tasks. The objective of the proposed mechanism is to minimize the aggregate delay of all tasks. A hybrid computation task offloading mechanism is proposed in [12], where the users' devices form the fog network. Each device may offload its computation tasks to nearby devices through device–to–device communication. The proposed mechanism aims to minimize the overall cost of all users using a centralized graph matching technique. A framework to enable developers and users to manage an IoT infrastructure is developed in [13]. This framework realizes the applications of IoT in smart cities. In [14], we also studied the computation task offloading in mobile cloud computing systems where only the remote cloud computing services are available. We proposed a joint optimal pricing and task scheduling algorithm for mobile devices with the objective of maximizing the utility of the users and the profit of the cloud service operator.

Similar to fog computing, mobile edge computing (MEC) enables the computing services at the edge of wireless cellular networks and is promising for the fifth generation (5G) wire-less systems [15]. In MEC, the computing servers are located within the radio access networks of cellular systems. Hence, unlike fog computing, the computation offloading is managed and controlled by the network operators. Computation task offloading in MEC has also been widely studied in recent years. Optimal offloading decisions of computation tasks are obtained in [16] by jointly allocating communication and computing resources. The objective is to minimize the overall cost of energy, computation, and delay for all mobile users. An efficient multi–user computation offloading mechanism is proposed in [17] that aims to minimize the cost of each mobile user individually. The mechanism allocates a wireless channel to each offloaded computation task assuming that there are sufficient computing resources at the network edge. The offloading decisions are made by a centralized controller located in the cellular base station.

### B. Motivation and Contributions

The existing works [7]–[12] studied the computation task offloading in fog computing by considering different objectives. However, they focused on optimizing the overall performance of the system (e.g., minimzing the total cost of the IoT system as in [7], [12] or minimizing the aggregate delay as in [8]–[11]). Nevertheless, selfish IoT users are interested in optimizing their own quality of experience (QoE) individually, which reflects their level of satisfactions of using computing services. They may not follow the strategies that aim to optimize the overall system performance and compete against each other for the limited fog computing resources. Thus, it is important to consider this competition in order to enable fog computing in real–world IoT systems.

In this paper, we propose an allocation mechanism for fog computing resources in IoT systems. Our goal is to determine the offloading decision for each task arriving to the IoT users, where each user is interested in maximizing its own QoE. Each user equipped with multi-RAT is able to offload its tasks to either different fog nodes or remote cloud servers. The limited processing power of each fog node results in competition among the IoT users when they intend to offload their tasks to a fog node. We first use a processor sharing method to allocate the computing resources of fog nodes and manage their limited resources. We then adopt a game theoretic approach to model the competition among IoT users.

In summary, the key contributions of this paper are as follows:

- *QoE maximization framework*: We formulate a QoE maximization problem for each user to determine its computation offloading decision. The QoE is the reduction in the computation energy and delay obtained by offloading the task to the fog nodes or remote cloud servers. Hence, it reflects the satisfaction of using computing services.
- *Game formulation*: We model the competition among IoT users as a potential game to determine the computation offloading decisions of all users. We analyze the properties of the formulated game and show the existence of a pure Nash equilibrium (NE). We further prove that the NE can be obtained in finite time. However, the time complexity to reach the NE may increase exponentially with the number of IoT users.
- *Equilibrium efficiency*: We prove that the price of anarchy (PoA), which reflects the equilibrium efficiency loss, is bounded. Thus, the degradation of the social cost due to the strategic behavior of players is no worse than a constant. We also investigate the efficiency of the equilibrium through numerical experiments and show that the proposed algorithm is able to achieve a close-to-optimal social cost.
- *Near–optimal resource allocation*: To address the time complexity of determining the equilibrium, we propose a near–optimal resource allocation algorithm. We also prove that for an IoT system with $N$ users, the proposed algorithm can achieve an $\epsilon$-Nash equilibrium in $O(N/\epsilon)$ steps, which is polynomial in $N$.
- *Performance evaluation*: We investigate the performance of the proposed resource allocation algorithm through extensive numerical experiments. We first study the users' QoE obtained at the equilibrium. Our results show that by utilizing the proposed algorithm, the IoT users can obtain a higher QoE. In particular, for delay-sensitive applications, the existence of fog nodes in the close proximity of users reduces the computation time by up to 70%. We further compare our algorithm with an existing job dispatching and allocation mechanism proposed in [8]. Results show that up to 20% more IoT users benefit from computing services when our proposed algorithm is used in comparison to that of [8].

This paper is organized as follows. In Section II, we introduce the system model. We formulate the potential game in Section III, prove the existence of an equilibrium, and develop an algorithm that can achieve the NE. We also provide an upper bound for the PoA. In Section IV, we extend our framework and propose a near–optimal resource allocation mechanism and prove that it achieves an $\epsilon$-Nash equilibrium. We evaluate the performance of our framework through extensive simulations in Section V. Finally, we conclude in Section VI.

## II. System Model

We consider an IoT system with a hierarchical computing structure and a set of IoT users. Each IoT user may either perform its tasks locally or offload them to the computing servers. The computing servers include a set of fog nodes and the remote cloud servers. Fog nodes can provide computing services to the IoT users in their close proximity. Fig. 1 illustrates an instant of such IoT system.

### A. Hierarchical Computing Structure

We assume that there are $S$ fog nodes in the system where $\{1, \ldots, S\}$ denotes the set of these fog nodes. We further denote the set of all computing servers available in the hierarchical computing structure as $\mathcal{S} = \{0, 1, \ldots, S\}$, where 0 represents the remote cloud servers and is used to model all remote cloud servers that IoT users can use for their computation task offloading. We assume that the remote cloud servers have sufficient computing resources. We model each cloud server as a virtual machine with dedicated processing power $f_0$, which denotes the processing capacity of the server in cycles per time unit. Without loss of generality, we assume that all remote cloud servers have the same processing power.

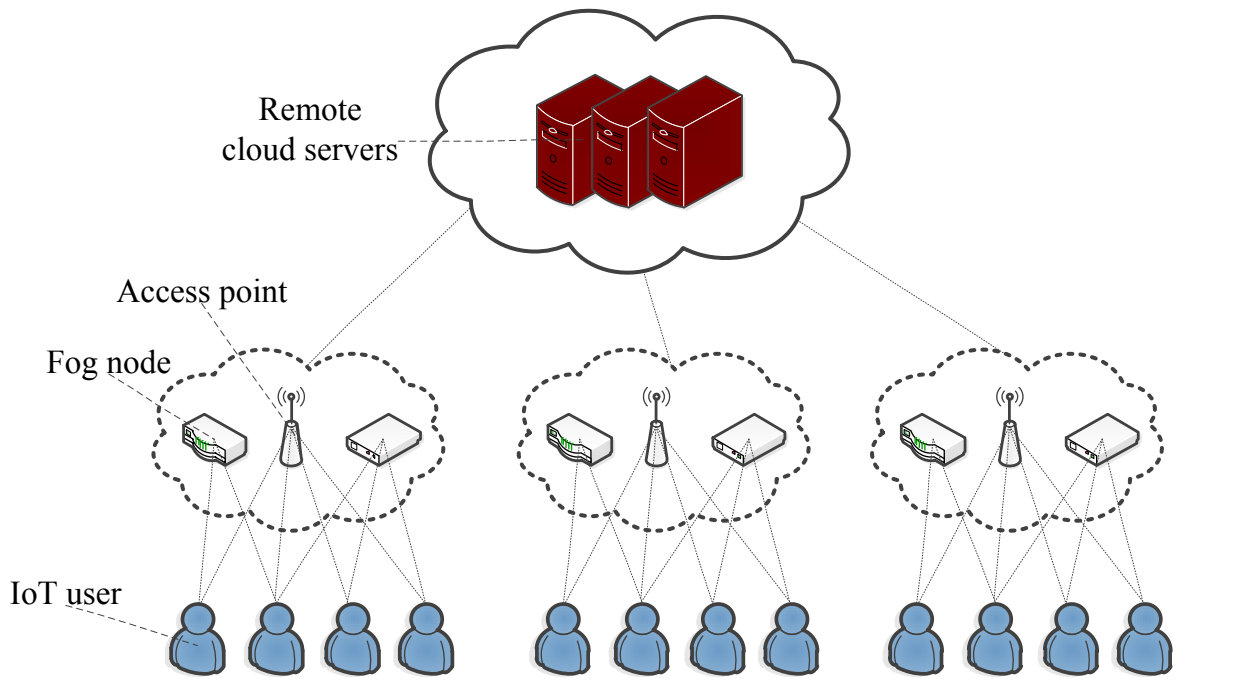

Fig. 1. An IoT system consists of IoT users, fog nodes, and the remote cloud servers. The IoT users may offload their computation tasks to either fog nodes which are located in close proximity or remote cloud servers via access points.

Unlike the remote cloud servers, fog nodes have limited processing power. We model the computing resource of each fog node $s = 1, \ldots, S$ as a virtual machine with processing power $f_s$. The processing power of each fog node is equally shared among the applications of IoT users offloaded to the fog node. Notice that the fog nodes cannot prioritize applications of an IoT user as all users are selfish and only follow the strategic behaviors that maximize their own QoE.

### B. IoT Users and Computation Task Models

We consider that there are $N$ IoT users denoted by the set $\mathcal{N} = \{1, \ldots, N\}$. Each IoT user $n \in \mathcal{N}$ has a computation task $\mathcal{T}_n = (z_n, \gamma_n)$, where $z_n$ is the size of task in bits. Furthermore, $\gamma_n$ is the processing density in cycles per bit, which is the number of cycles required to process a unit bit of data. The processing density of each task depends on the application type and is known to the user upon arrival of the task. Each IoT user may perform its task locally or offload it to the computing servers. The computing server can either be one of the fog nodes in close proximity or belong to the remote cloud. Although each IoT user is equipped with multi-RAT and may have access to more than one computing server, each computation task should either be served entirely by the IoT device of the user or be offloaded to one computing server. We define the offloading indicator $a_{n,s} \in \{0, 1\}$ for each IoT user $n \in \mathcal{N}$ and computing server $s \in \mathcal{S}$, where $a_{n,s} = 1$ indicates that the task $\mathcal{T}_n$ from IoT user $n$ is offloaded to the server $s$. Otherwise, if $a_{n,s} = 0$ for all $s \in \mathcal{S}$, then task $\mathcal{T}_n$ is performed locally by the user. We also define the *offloading vector* of user $n \in \mathcal{N}$ as $\mathbf{a}_n = (a_{n,0}, \ldots, a_{n,S})$. Since each task $n$ can be offloaded to at most one server, we have $\sum_{s \in \mathcal{S}} a_{n,s} \leq 1$. We further introduce constant $b_{n,s} \in \{0, 1\}$ for each $n \in \mathcal{N}$ and $s \in \mathcal{S}$, which represents the connectivity of user $n$ to computing server $s$. If $b_{n,s} = 1$, then user $n$ is able to offload its task to the server $s$. Otherwise, if user $n$ is not in close proximity of fog node $s$ and cannot connect to it, we have $b_{n,s} = 0$ to exclude $s$ from the feasible strategy space of user $n$. We restrict the offloading indicator to take the connectivity model into account and ensure that $a_{n,s} \leq b_{n,s}$ for all $n \in \mathcal{N}$ and $s \in \mathcal{S}$.

We now introduce the QoE of IoT users when performing the computation tasks. The QoE reflects the benefits an IoT user received by offloading its tasks to the computing servers. We define the QoE as the cost reduction achieved from computation offloading, where the cost of performing a task consists of the computation energy and computation delay. We first determine the cost of performing a task in both local and fog–cloud computing cases.

*Case 1. Local Computing:* The computation energy for performing a task of size $z_n$ and processing density $\gamma_n$ locally in IoT device $n$ is $\alpha_n z_n \gamma_n / f_n$, where $f_n$ denotes the processing power of IoT device $n$ and $\alpha_n$ is a user–dependent constant. The same model of energy consumption is used in [18]–[20]. Furthermore, the time required to process the task is $z_n \gamma_n / f_n$. We define the cost of performing the task $\mathcal{T}_n = (z_n, \gamma_n)$ locally as follows:

$$c_n^{\mathrm{L}} \triangleq \lambda_n^{\mathrm{E}} \alpha_n \frac{z_n \gamma_n}{f_n} + \lambda_n^{\mathrm{T}} \frac{z_n \gamma_n}{f_n}, \tag{1}$$

where $\lambda_n^{\mathrm{E}}, \lambda_n^{\mathrm{T}}$ denote the constant weights of computation energy and computation time, respectively. We assume that $\lambda_n^{\mathrm{E}} \in [0, 1]$, while $\lambda_n^{\mathrm{T}} \in (0, 1]$ to avoid the case that a task experiences a huge delay. These parameters depend on the IoT users and the type of application.

*Case 2. Fog–Cloud Computing:* In this case, an IoT user offloads its computation task to either a fog node in close proximity or the remote cloud server. The task will be submitted to the computing server via the user's wireless interface. We assume that for each user $n$ and computing server $s$, at most one wireless interface is used as indicated by constant $b_{n,s}$. For IoT user $n$, we denote the energy required to offload the task to computing server $s$ as $\beta_{n,s} z_n / r_{n,s}$, where $r_{n,s}$ is the data rate of the wireless interface transmitting the task to the computing server $s$ and constant $\beta_{n,s}$ depends on the type of the wireless interface. Similar to [8], [17], we assume that the transmission rate can be obtained by measurement and is known to the user. For an offloaded task, the delay consists of the time to dispatch the task to the allocated computing server and the computation time required to complete the task on the server. The former includes the transmission time of the wireless interface, which is $z_n / r_{n,s}$, and the roundtrip delay between the IoT device and the server. The roundtrip delay to access fog nodes is negligible as they are located in close proximity of the IoT users, while dispatching the tasks to the remote cloud servers may impose a long roundtrip delay. We denote the roundtrip delay between IoT device $n$ and computing server $s$ as $d_{n,s}^{\mathrm{rt}}$.

We now determine the computation time each offloaded task spends in the computing servers. Let matrix $\mathbf{A}_{-n}$ denote the offloading strategies of all IoT users excluding user $n$. Each row of $\mathbf{A}_{-n}$ represents the offloading vector of a user. We define $d_{n,s}^{\mathrm{c}}(\mathbf{A}_{-n})$ as the computation time required to process task $\mathcal{T}_n$ in server $s$, when the task is being offloaded. Similar to other existing works (e.g., [7], [9], [16], [17]), we focus on a particular time instant and allocate the computing services to the jobs arrived at the system.[1] We further assume that the processing power of fog nodes is equally shared among the workload that is arrived from IoT users and is under processing in the fog nodes. We now determine the computation time each task $\mathcal{T}_n$ spends in fog nodes. For fog node $s \in \mathcal{S}\setminus\{0\}$ and task $\mathcal{T}_n, n \in \mathcal{N}$, if it is the only task in the server (i.e., $a_{m,s} = 0$ for all $m \in \mathcal{N} \setminus \{n\}$), the processing power of the fog node is assigned to this task and it takes $d^{\mathrm{c}}_{n,s} = z_n\gamma_n/f_s$ to process the task. The computation time may increase when other IoT users are allocated to this server. If there are more tasks other than $\mathcal{T}_n$ arrived at the fog node, the processing power of the fog node is equally shared among them. The number of these tasks including task $\mathcal{T}_n$ is $1+\sum_{m\in\mathcal{N}\setminus\{n\}} a_{m,s}$. Let assume task $\mathcal{T}_n$ is the smallest task, i.e., $z_n\gamma_n < z_m\gamma_m$ for all $m \in \mathcal{N} \setminus \{n\}$. In this case, the computation time required to process this task is:

$$\frac{z_n\gamma_n}{f_s}\left(1+\sum_{m\in\mathcal{N}\setminus\{n\}} a_{m,s}\right),$$

as the processing power of the fog node is equally shared among the offloaded tasks. However, smaller tasks depart the server sooner as they require less processing power. Once a task leaves the server, we reassign the processing power to other tasks which are still under processing. Therefore, the computation time of task $\mathcal{T}_n$ offloaded to fog node $s \in \mathcal{S}\setminus\{0\}$ is

$$d^{\mathrm{c}}_{n,s}\left(\mathbf{A}_{-n}\right) = \frac{z_n\gamma_n}{f_s}\left(1+\sum_{m\in\mathcal{N}\setminus\{n\}} \min\left\{\frac{z_m\gamma_m}{z_n\gamma_n}, 1\right\} a_{m,s}\right). \tag{2}$$

Notice that $\min\{z_m\gamma_m/z_n\gamma_n, 1\} = 1$ when $z_m\gamma_m > z_n\gamma_n$. Otherwise, this term is equal to $z_m\gamma_m/z_n\gamma_n < 1$, which states that task $\mathcal{T}_m$ leaves the fog node earlier than task $\mathcal{T}_n$. Given $\mathbf{A}_{-n}$ as the offloading strategies of IoT users except user $n$, the second term of $d^{\mathrm{c}}_{n,s}\left(\mathbf{A}_{-n}\right)$ determines how the other tasks offloaded to server $s$ prolong the computation time of task $\mathcal{T}_n$.

The computation time spent in the remote cloud servers can also be determined as follows as we assume that they have sufficient processing power.

$$d^{\mathrm{c}}_{n,0}\left(\mathbf{A}_{-n}\right) = \frac{z_n\gamma_n}{f_0}. \tag{3}$$

The cost imposed to IoT user $n$ when offloading the task $\mathcal{T}_n$ is:

$$\begin{aligned} c^{\mathrm{C}}_n\left(\mathbf{a}_n, \mathbf{A}_{-n}\right) \triangleq\ & \lambda^{\mathrm{E}}_n \sum_{s\in\mathcal{S}} a_{n,s}\frac{\beta_{n,s}z_n}{r_{n,s}} \\ & + \lambda^{\mathrm{T}}_n \sum_{s\in\mathcal{S}} a_{n,s}\left(\frac{z_n}{r_{n,s}} + d^{\mathrm{rt}}_{n,s} + d^{\mathrm{c}}_{n,s}\left(\mathbf{A}_{-n}\right)\right). \end{aligned} \tag{4}$$

We further define the QoE as the amount of cost reduction achieved by the user when offloading its task. Let $q_n\left(\mathbf{a}_n, \mathbf{A}_{-n}\right)$ denote the user $n$'s QoE given offloading vector $\mathbf{a}_n$ and matrix $\mathbf{A}_{-n}$. We have

$$q_n\left(\mathbf{a}_n, \mathbf{A}_{-n}\right) = \begin{cases} c^{\mathrm{L}}_n - c^{\mathrm{C}}_n\left(\mathbf{a}_n, \mathbf{A}_{-n}\right), & \text{if } \sum_{s\in\mathcal{S}} a_{n,s} = 1 \\ 0, & \text{otherwise,} \end{cases} \tag{5}$$

where $c^{\mathrm{L}}_n$ and $c^{\mathrm{C}}_n\left(\mathbf{a}_n, \mathbf{A}_{-n}\right)$ are given in (1) and (4), respectively. Note that when $\sum_{s\in\mathcal{S}} a_{n,s} = 1$, the task is offloaded and the user's QoE is the cost of local computing minus the cost imposed by offloading the task. Otherwise, the QoE is zero as the task is performed locally.

## III. Computation Offloading Game

In this section, we formulate the interactions between the IoT users as a strategic game and propose an algorithm that can obtain the NE. We further analyze the PoA for this game.

### A. Game Formulation

We formally define game $\mathcal{G} \triangleq \left(\mathcal{N}, \prod_{n\in\mathcal{N}} \mathcal{A}_n, \{q_n\}_{n\in\mathcal{N}}\right)$, where $\mathcal{N}$ is the set of players and $\mathcal{A}_n$ is the feasible strategy space of player $n$ such that $\mathbf{a}_n \in \mathcal{A}_n$. Furthermore, $q_n$ is the player $n$'s QoE, which represents its payoff achieved from using computing services.

Each IoT user aims to maximize its own QoE in response to the other users' strategies. To obtain the strategies of all IoT users, we first introduce the concept of best response strategy.

**Definition 1** (**Best Response Strategy** [21]). *Given $\mathbf{A}_{-n}$ as the strategies of all players excluding player $n$, player $n$'s best response strategy is:*

$$\mathbf{a}^*_n = \underset{\mathbf{a}_n}{\operatorname{argmax}} \quad q_n\left(\mathbf{a}_n, \mathbf{A}_{-n}\right) \tag{6a}$$

$$\text{subject to } \sum_{s\in\mathcal{S}} a_{n,s} \leq 1, \tag{6b}$$

$$a_{n,s} \leq b_{n,s}, \quad \forall s \in \mathcal{S}, \tag{6c}$$

$$a_{n,s} \in \{0, 1\}, \quad \forall s \in \mathcal{S}, \tag{6d}$$

*which represents the choice of $\mathbf{a}_n$ that maximizes the player $n$'s QoE.*

In QoE maximization problem (6), constraint (6b) is introduced to guarantee that each task can be offloaded to at most one computing server. Constraint (6c) ensures that offloading to computing server $s$ is possible if there is a wireless link connecting user $n$ to server $s$. The feasible region of problem (6) represents the feasible strategy space of player $n$, which is denoted as $\mathcal{A}_n$.

Utilizing Definition 1, we now introduce the NE as follows.

**Definition 2** (**Nash Equilibrium** [21]). *An offloading strategy profile $\{\mathbf{a}^*_n\}_{n\in\mathcal{N}}$ is an NE of Game $\mathcal{G}$ if it is a fixed point of best responses, i.e., for all $\mathbf{a}'_n \in \mathcal{A}_n, n \in \mathcal{N}$*

$$q_n\left(\mathbf{a}^*_n, \mathbf{A}^*_{-n}\right) \geq q_n\left(\mathbf{a}'_n, \mathbf{A}^*_{-n}\right).$$

According to Definition 2, no player has an incentive to deviate unilaterally from the NE as the player cannot further improve its QoE by following a different strategy than the equilibrium.

[1] An online mechanism design can address the arbitrary arrival of jobs and we will leave this extension as future work.

We now show that there exists an NE for Game $\mathcal{G}$. To obtain the NE, we use the potential games [22], [23] and introduce the weighted potential game as follows.

**Definition 3 (Weighted Potential Game).** *Let* $\mathbf{w} = (w_n)_{n\in\mathcal{N}}$ *denote a vector of positive numbers. A game is called a weighted potential game if it admits a* $w$*–potential function* $P$ *such that for every player* $n \in \mathcal{N}$ *and offloading vectors* $\mathbf{a}_n, \mathbf{a}'_n \in \mathcal{A}_n$*,*

$$q_n(\mathbf{a}_n, \mathbf{A}_{-n}) - q_n(\mathbf{a}'_n, \mathbf{A}_{-n}) = w_n\left(P(\mathbf{a}_n, \mathbf{A}_{-n}) - P(\mathbf{a}'_n, \mathbf{A}_{-n})\right).$$

In a weighted potential game, each player $n$ is associated with a positive weight as denoted by $w_n$. To formulate the potential game, we first define the function $Q(\mathbf{A})$ as the weighted aggregate QoE of all users.

$$Q(\mathbf{A}) \triangleq \sum_{n\in\mathcal{N}} \frac{1}{\lambda_n^{\mathrm{T}}} q_n(\mathbf{a}_n, \mathbf{A}_{-n}), \tag{7}$$

where $\mathbf{A} = (\mathbf{a}_n, \mathbf{A}_{-n})$. We further define the function $\bar{Q}(\mathbf{A})$ as the weighted aggregate QoE of all users if each user is alone in the game.

$$\bar{Q}(\mathbf{A}) \triangleq \sum_{n\in\mathcal{N}} \frac{1}{\lambda_n^{\mathrm{T}}} q_n(\mathbf{a}_n, \mathbf{0}), \tag{8}$$

where $\mathbf{0} = (0)_{(N-1)\times|\mathcal{S}|}$ is an all zero $(N-1) \times |\mathcal{S}|$ matrix. In the following theorem, we introduce a $w$–potential function and show that it satisfies the condition of Definition 3.

**Theorem 1.** *Given vector* $\mathbf{w} = \left(\lambda_n^T\right)_{n\in\mathcal{N}}$*, the following function is a* $w$*–potential function and game* $\mathcal{G}$ *is a weighted potential game.*

$$P(\mathbf{A}) \triangleq \frac{Q(\mathbf{A}) + \bar{Q}(\mathbf{A})}{2}. \tag{9}$$

*Proof.* We prove Theorem 1 when we show that the potential function introduced in this theorem is a $w$–weighted potential function satisfying the condition of Definition 3. To facilitate the analysis, we define $\rho_{n,s,m} \triangleq \frac{z_n\gamma_n}{f_s}\min\left\{\frac{z_m\gamma_m}{z_n\gamma_n}, 1\right\}$. We first form

$$\begin{aligned}
P(\mathbf{a}_n,\, &\mathbf{A}_{-n}) - P(\mathbf{a}'_n, \mathbf{A}_{-n}) \\
&= \frac{Q(\mathbf{a}_n, \mathbf{A}_{-n}) - Q(\mathbf{a}'_n, \mathbf{A}_{-n})}{2} + \frac{\bar{Q}(\mathbf{a}_n, \mathbf{0}) - \bar{Q}(\mathbf{a}'_n, \mathbf{0})}{2} \\
&= \sum_{n\in\mathcal{N}} \frac{1}{2\lambda_n^{\mathrm{T}}}\left(q_n(\mathbf{a}_n, \mathbf{A}_{-n}) - q_n(\mathbf{a}'_n, \mathbf{A}_{-n})\right) \\
&\quad + \frac{1}{2\lambda_n^{\mathrm{T}}}\left(q_n(\mathbf{a}_n, \mathbf{0}) - q_n(\mathbf{a}'_n, \mathbf{0})\right) \\
&= -\sum_{s\in\mathcal{S}}(a_{n,s} - a'_{n,s})\frac{\lambda_n^{\mathrm{E}}}{\lambda_n^{\mathrm{T}}}\frac{\beta_{n,s}z_n}{r_{n,s}} \\
&\quad -\sum_{s\in\mathcal{S}}(a_{n,s} - a'_{n,s})\left(\frac{z_n\gamma_n}{f_s} + \frac{z_n}{r_{n,s}} + d_{n,s}^{\mathrm{rt}}\right) \\
&\quad -\frac{1}{2}\sum_{m\in\mathcal{N}\setminus\{n\}}\sum_{s\in\mathcal{S}}(a_{n,s} - a'_{n,s})a_{m,s}\rho_{n,s,m} \\
&\quad -\frac{1}{2}\sum_{m\in\mathcal{N}\setminus\{n\}}\sum_{s\in\mathcal{S}}a_{m,s}(a_{n,s} - a'_{n,s})\rho_{m,s,n} \\
&\overset{(a)}{=} -\frac{\lambda_n^{\mathrm{E}}}{\lambda_n^{\mathrm{T}}}\sum_{s\in\mathcal{S}}(a_{n,s} - a'_{n,s})\frac{\beta_{n,s}z_n}{r_{n,s}} \\
&\quad -\sum_{s\in\mathcal{S}}(a_{n,s} - a'_{n,s})\left(\frac{z_n}{r_{n,s}} + d_{n,s}^{\mathrm{rt}} + d_{n,s}^{\mathrm{c}}(\mathbf{A}_{-n})\right) \\
&= \frac{1}{\lambda_n^{\mathrm{T}}}\left(q_n(\mathbf{a}_n, \mathbf{A}_{-n}) - q_n(\mathbf{a}'_n, \mathbf{A}_{-n})\right),
\end{aligned}$$

where $(a)$ is obtained based on the fact that $\rho_{n,s,m} = \rho_{m,s,n}$ for all $n, m \in \mathcal{N}$ and $s \in \mathcal{S}$. Notice that

$$\frac{z_n\gamma_n}{f_s}\min\left\{\frac{z_m\gamma_m}{z_n\gamma_n}, 1\right\} = \frac{z_m\gamma_m}{f_s}\min\left\{\frac{z_n\gamma_n}{z_m\gamma_m}, 1\right\}.$$

Therefore, the introduced function is a $w$–potential function with weights $w_n = \lambda_n^{\mathrm{T}}$ for all $n \in \mathcal{N}$. Accordingly, we conclude that Game $\mathcal{G}$ is a weighted potential game. ■

---

**Algorithm 1:** Best Response Adaptation for an IoT user $n$.

---

**initialization**: $t \leftarrow 0$, and $\mathbf{a}_n^{*(0)}$
**do**
  $t \leftarrow t+1$
  User $n$ collects $d_{n,s}(\mathbf{A}_{-n})$ from fog nodes.
  User $n$ updates its best response strategy

$$\mathbf{a}_n^{*(t)} \leftarrow \underset{\mathbf{a}_n}{\operatorname{argmax}}\, q_n(\mathbf{a}_n, \mathbf{A}_{-n}) \quad \text{subject to } (6b) - (6d),$$

  and submits it to the neighboring fog nodes.
**while** $\mathbf{a}_n^{*(t)} \neq \mathbf{a}_n^{*(t-1)}$
**output**: $\mathbf{a}_n^{\mathrm{NE}} \leftarrow \mathbf{a}_n^{*(t)}$

---

In the following lemma, we now show that there exists a pure NE in Game $G$.

**Lemma 1.** *Every finite potential game possesses a pure-strategy NE and has the finite improvement property.*

The proof of Lemma 1 can be found in [22]. Lemma 1 states that there exists an NE for every potential game with finite strategy space of players. This further implies that any algorithm that updates the players' strategies and improves their QoE is guaranteed to reach an NE in finite time. Utilizing this lemma, we now propose a best response algorithm as illustrated in Algorithm 1.

In Algorithm 1, we update the best response strategy of IoT user $n$ in an iterative manner. In each iteration, the user determines its best strategy in response to other users' strategies by solving problem (6). Note that user $n$ only needs to know the value of computation delay $d_{n,s}(\mathbf{A}_{-n})$ for fog nodes $s$ in close proximity. Each fog node can measure this delay and reports to the IoT users in its close proximity. According to Definition 2, the fixed point of best response strategies of all users is the NE of Game $\mathcal{G}$ if the best response algorithm converges. Let $\mathbf{a}_n^{\mathrm{NE}}$ denote the equilibrium strategy of user $n$. Algorithm 1 converges to $\mathbf{a}_n^{\mathrm{NE}}$ in finite time as stated in Lemma 1. However, the time of convergence increases exponentially with the number of IoT users $N$ and is $O(2^{NS})$ in the worst case.

### B. Price of Anarchy (PoA)

We have so far shown that Game $\mathcal{G}$ possesses at least an NE, which can be obtained using Algorithm 1. We now study an important performance metric of strategic games to answer the following question. *How far is the overall performance of an NE from the socially optimal allocation?* We use the PoA to quantify this difference. In a strategic game, the PoA illustrates how the social cost degrades due to players' selfish behaviors. To facilitate the analysis of PoA, we resort to the IoT users' cost minimization framework as it admits the same equilibrium as the QoE maximization framework. We define the cost of user $n \in \mathcal{N}$ as

$$c_n\left(\mathbf{a}_n, \mathbf{A}_{-n}\right) = \begin{cases} c_n^{\mathrm{C}}\left(\mathbf{a}_n, \mathbf{A}_{-n}\right), & \text{if } \sum_{s\in\mathcal{S}} a_{n,s} = 1 \\ c_n^{\mathrm{L}}, & \text{otherwise.} \end{cases} \tag{10}$$

The PoA is defined as the ratio between the worst social cost obtained in an NE and the optimal centralized solution of the social cost minimization problem. In a strategic game, the PoA illustrates how the social cost increases due to players' selfish behaviors.

The social cost is defined as the aggregate cost of all IoT users as follows.

$$\Phi\left(\mathbf{A}\right) \triangleq \sum_{n\in\mathcal{N}} c_n\left(\mathbf{a}_n, \mathbf{A}_{-n}\right). \tag{11}$$

The PoA can then be obtained as

$$PoA = \frac{\max_{\mathbf{A}\in\mathcal{A}^{NE}} \Phi\left(\mathbf{A}\right)}{\underset{\mathbf{A}\in\mathcal{A}}{\text{minimize}}\ \Phi\left(\mathbf{A}\right)}, \tag{12}$$

where $\mathcal{A} = \prod_{n\in\mathcal{N}} \mathcal{A}_n$ is the strategy space of all users and $\mathcal{A}^{\mathrm{NE}}$ is the set of all equilibria.

In the following theorem, we provide an upper bound for the PoA of the NE in Game $\mathcal{G}$ and show that it is no worse than a constant.

**Theorem 2.** *The PoA of the computation offloading game $\mathcal{G}$ is no worse than*

$$\min\left\{N_s, \frac{\sum_{n\in\mathcal{N}} c_n^L}{\sum_{n\in\mathcal{N}} \min_{\mathbf{a}_n\in\mathcal{A}_n} c_n\left(\mathbf{a}_n, \mathbf{0}\right)}\right\},$$

*where $N_s$ denotes the maximum number of IoT users allocated to a fog node and $\mathbf{0}$ is an all zero $(N-1)\times|\mathcal{S}|$ matrix.*

*Proof.* We first show that PoA is no larger than $N_s$, which is the maximum number of IoT users allocated to a fog node. Notice that $N_s$ is much less than the number of users and hence it is a tight bound for the PoA. From Definition 2, we know that for any user $n \in \mathcal{N}$ at the equilibrium, we have

$$c_n\left(\mathbf{a}_n^{\mathrm{NE}}, \mathbf{A}_{-n}^{\mathrm{NE}}\right) \le c_n\left(\mathbf{a}_n^{\mathrm{S}}, \mathbf{A}_{-n}^{\mathrm{NE}}\right), \tag{13}$$

where $\mathbf{a}_n^{\mathrm{S}}$ denotes the socially optimal strategy of user $n$. For the user $n$'s strategy, we consider two cases of local task execution and offloading. For the case of task offloading, we have

$$\begin{aligned} c_n\left(\mathbf{a}_n^{\mathrm{S}}, \mathbf{A}_{-n}^{\mathrm{NE}}\right) &= c_n^{\mathrm{C}}\left(\mathbf{a}_n^{\mathrm{S}}, \mathbf{A}_{-n}^{\mathrm{NE}}\right) \\ &= \lambda_n^{\mathrm{E}} \sum_{s\in\mathcal{S}} a_{n,s}^{\mathrm{S}} \frac{\beta_{n,s} z_n}{r_{n,s}} \\ &\quad + \lambda_n^{\mathrm{T}} \sum_{s\in\mathcal{S}} a_{n,s}^{\mathrm{S}} \left(\frac{z_n}{r_{n,s}} + d_{n,s}^{\mathrm{rt}}\right) \\ &\quad + \lambda_n^{\mathrm{T}} \sum_{s\in\mathcal{S}} a_{n,s}^{\mathrm{S}} d_{n,s}^{\mathrm{c}}\left(\mathbf{A}_{-n}^{\mathrm{NE}}\right). \end{aligned}$$

Notice that in $c_n^{\mathrm{C}}\left(\mathbf{a}_n^{\mathrm{S}}, \mathbf{A}_{-n}^{\mathrm{NE}}\right)$, only the last term depends on the strategies of other players (i.e., $\mathbf{A}_{-n}^{\mathrm{NE}}$). If other users change their strategies from $\mathbf{A}_{-n}^{\mathrm{NE}}$ to the socially optimal strategy $\mathbf{A}_{-n}^{\mathrm{S}}$, $d_{n,s}^{\mathrm{c}}\left(\mathbf{A}_{-n}^{\mathrm{NE}}\right)$ will be increased by at most $N_s$ times. Therefore, for the last term of $c_n^{\mathrm{C}}\left(\mathbf{a}_n^{\mathrm{S}}, \mathbf{A}_{-n}^{\mathrm{NE}}\right)$, we have

$$\sum_{s\in\mathcal{S}} a_{n,s}^{\mathrm{S}} d_{n,s}^{\mathrm{c}}\left(\mathbf{A}_{-n}^{\mathrm{NE}}\right) \le N_s \sum_{s\in\mathcal{S}} a_{n,s}^{\mathrm{S}} d_{n,s}^{\mathrm{c}}\left(\mathbf{A}_{-n}^{\mathrm{S}}\right). \tag{14}$$

Hence, in this case,

$$c_n\left(\mathbf{a}_n^{\mathrm{S}}, \mathbf{A}_{-n}^{\mathrm{NE}}\right) \le N_s c_n\left(\mathbf{a}_n^{\mathrm{S}}, \mathbf{A}_{-n}^{\mathrm{S}}\right).$$

We also know that the above inequality holds for the case of local computing as $c_n^{\mathrm{L}} \le N_s c_n^{\mathrm{L}}$. Thus, for any NE strategy including the worst equilibrium, we have

$$c_n\left(\mathbf{a}_n^{\mathrm{NE}}, \mathbf{A}_{-n}^{\mathrm{NE}}\right) \le N_s c_n\left(\mathbf{a}_n^{\mathrm{S}}, \mathbf{A}_{-n}^{\mathrm{S}}\right), \quad \forall n \in \mathcal{N}.$$

We now conclude that PoA is always less than or equal to $N_s$.

We further show that the PoA is also no larger than the second term presented in Theorem 2. We show that if all users choose local task execution in an NE, that equilibrium is the worst one. In other words, that equilibrium results in a higher social cost than all other equilibria. In this case, the maximum social cost is $\sum_{n\in\mathcal{N}} c_n^{\mathrm{L}}$. By contradiction, we assume that there exists an NE in which the social cost is higher than $\sum_{n\in\mathcal{N}} c_n^{\mathrm{L}}$. Therefore, there should be at least one user $n$ such that $c_n\left(\mathbf{a}_n^{\mathrm{NE}}, \mathbf{A}_{-n}^{\mathrm{NE}}\right) > c_n^{\mathrm{L}}$. However, according to Definition 2, this user has a positive incentive to move from the equilibrium and choose the local execution as it imposes less cost. Thus, if there is an NE in which all users choose the local execution, it is the worst equilibrium and we can conclude that

$$\max_{\mathbf{A}\in\mathcal{A}^{NE}} \Phi\left(\mathbf{A}\right) \le \sum_{n\in\mathcal{N}} c_n^{\mathrm{L}}.$$

We now focus on the minimum social cost and derive a lower bound. Let $\mathbf{a}_n^{\mathrm{S}}$ denote the strategy of user $n$ when minimizing the social cost. A lower bound of the cost of each user can be obtained when there is no competition with other users. Therefore,

$$\begin{aligned} c_n\left(\mathbf{a}_n^{\mathrm{S}}, \mathbf{A}_{-n}^{\mathrm{S}}\right) &\ge c_n\left(\mathbf{a}_n^{\mathrm{S}}, \mathbf{0}\right) \\ &\ge \min_{\mathbf{a}_n\in\mathcal{A}_n} c_n\left(\mathbf{a}_n, \mathbf{0}\right), \end{aligned}$$

which completes the proof. ■

**Algorithm 2:** $\epsilon$-Better Response Adaptation for an IoT user $n$.

**input**: $\epsilon$
**initialization**: $t \leftarrow 0$, $\mathcal{B}_n \leftarrow \mathcal{A}_n$, and $\mathbf{a}_n^{(0)}$
**while** $\mathcal{B}_n \neq \emptyset$
  User $n$ submits its better response strategy $\mathbf{a}_n^{(t)}$ to the neighboring fog nodes.
  User $n$ collects $d_{n,s}(\mathbf{A}_{-n})$ from fog nodes.
  User $n$ obtains a set of better response strategies

$$\mathcal{B}_n = \left\{\mathbf{a}_n \in \mathcal{A}_n |\ q_n(\mathbf{a}_n, \mathbf{A}_{-n}) - q_n\left(\mathbf{a}_n^{(t)}, \mathbf{A}_{-n}\right) > \epsilon\right\}$$

  User $n$ chooses $\mathbf{a}_n^{(t+1)}$ from set $\mathcal{B}_n$ and updates its strategy
  $t \leftarrow t+1$
**end**
**output**: $\mathbf{a}_n^{\epsilon} \leftarrow \mathbf{a}_n^{(t)}$

## IV. Near–Optimal Resource Allocation Mechanism

To address the time complexity of the best response algorithm presented in Section III, in this section, we propose a near-optimal resource allocation algorithm that terminates in polynomial time. We develop the algorithm by using a better response approach and show that it can achieve a near NE solution and approximately satisfies the NE condition. We first define the concept of $\epsilon$-Nash equilibrium.

**Definition 4** (**$\epsilon$-Nash Equilibrium** [21]). *An offloading strategy profile $\{\mathbf{a}_n^{\epsilon}\}_{n\in\mathcal{N}}$ is an $\epsilon$-Nash equilibrium of Game $\mathcal{G}$ if for all $\mathbf{a}_n \in \mathcal{A}_n, n \in \mathcal{N}$*

$$q_n\left(\mathbf{a}_n^{\epsilon}, \mathbf{A}_{-n}^{\epsilon}\right) \geq q_n\left(\mathbf{a}_n, \mathbf{A}_{-n}^{\epsilon}\right) - \epsilon.$$

In an NE, no player has an incentive to change its strategy. However, in an $\epsilon$-Nash equilibrium, this requirement is weakened to allow the possibility that a player may have a small bounded incentive to deviate from NE. The player cannot expect to increase its payoff (i.e., QoE) by more than $\epsilon$.

We now develop an algorithm that can achieve an $\epsilon$-Nash equilibrium. In this algorithm, each player updates its strategy to a better response strategy rather than its best response strategy. Given a constant $\epsilon$, if there exists a strategy which improves the player's QoE by more than $\epsilon$, the player updates its strategy to one such strategy. Otherwise, the player does not change its strategy assuming that it has reached near an NE. Algorithm 2 illustrates our proposed better response algorithm for IoT user $n$.

In Algorithm 2, for each IoT user, we update its strategy to a better one which improves the user's QoE by more than $\epsilon$. Set $\mathcal{B}_n$ shows all feasible better strategies that user $n$ can follow. Once this set is empty, we terminate the algorithm as there is no such strategy. In the following theorem, we show that Algorithm 2 can achieve the $\epsilon$-Nash equilibrium with a polynomial time complexity.

**Theorem 3.** *For any given $\epsilon > 0$, the better response algorithm illustrated in Algorithm 2 reaches the $\epsilon$-Nash equilibrium in $O(N/\epsilon)$ steps.*

*Proof.* According to Definition 4, the equilibrium obtained by Algorithm 2 is an $\epsilon$-Nash equilibrium as the user incentive obtained by deviating from the equilibrium is at most $\epsilon$. We now prove that the time complexity of the algorithm is at most $O(N/\epsilon)$. In each iteration $t$, user $n$ improves its QoE by at least $\epsilon$. Thus,

$$q_n\left(\mathbf{a}_n^{(t)}, \mathbf{A}_{-n}\right) - q_n\left(\mathbf{a}_n^{(t-1)}, \mathbf{A}_{-n}\right) > \epsilon.$$

Since Game $\mathcal{G}$ is a weighted potential game as stated in Theorem 1, we have

$$P\left(\mathbf{a}_n^{(t)}, \mathbf{A}_{-n}\right) - P\left(\mathbf{a}_n^{(t-1)}, \mathbf{A}_{-n}\right) > \frac{\epsilon}{\lambda_n^{\mathrm{T}}}.$$

As a result, each IoT user increases the potential function $P$ by at least $\epsilon/\lambda_n^{\mathrm{T}}$ in each iteration. However, the potential function $P$ is bounded above and is always less than $\sum_{n\in\mathcal{N}} c_n^{\mathrm{L}}/\lambda_n^{\mathrm{T}}$. Therefore, the number of better response updates is at most

$$\frac{\Lambda^{\mathrm{T}}}{\epsilon}\sum_{n\in\mathcal{N}}\frac{c_n^{\mathrm{L}}}{\lambda_n^{\mathrm{T}}},$$

where $\Lambda^{\mathrm{T}} = \max_{n\in\mathcal{N}} \lambda_n^{\mathrm{T}}$. We can conclude that the time complexity of our proposed algorithm to determine the $\epsilon$-Nash equilibrium in all users is $O(N/\epsilon)$, which completes the proof. ■

## V. Performance Evaluation

In this section, we investigate the performance of the proposed algorithm by evaluating the IoT users' QoE at the equilibrium and that of socially optimum mechanism. We further compare our proposed algorithm with an existing algorithm proposed in [8].

### A. Simulation Setup

For each IoT device, we assume that the CPU clock speed is randomly and uniformly taken from [100 MHz, 1 GHz].[2] To determine the energy consumption of the CPU, we follow the model presented in [18]–[20]. For a task of size $z_n$ and processing density $\gamma_n$, the energy consumed in the device $n$'s CPU to perform the task is $\alpha_n z_n \gamma_n / f_n$. The user-dependent parameter $\alpha_n$ is [18]–[20]:

$$\alpha_n = \kappa_n f_n^{\varphi_n} + \varrho_n,$$

where $\kappa_n$, $\varphi_n$, and $\varrho_n$ are user-dependent constants that depend on the CPU model. We use the measurement results reported in [26] and set $\kappa_n = 0.33$, $\varphi_n = 3$, $\varrho_n = 0.1$ for all $n \in \mathcal{N}$. We assume that each IoT device is equipped with three wireless interfaces, which are Long Term Evolution (LTE), WiFi, and Bluetooth interfaces. IoT users may use their LTE interface to communicate with the remote cloud servers, while they can use the WiFi and Bluetooth interfaces to connect to at most two nearby fog nodes. Without loss of generality, we assume that the energy parameter $\beta_{n,s}$ only depends on the type of interface and does not vary in different devices. According to [26], for the LTE interface of each device $n \in \mathcal{N}$, we set $\beta_{n,0} = 2605$ mJ/sec. We further set $\beta_{n,s} = 1224.78$ mJ/sec [27] if user $n$ is connected to fog node $s$ via WiFi interface and $\beta_{n,s} = 84$ mJ/sec [28] if the Bluetooth interface is used. The average transmission rate of LTE and WiFi are 5.85 Mbps and 3.01 Mbps, respectively, as measured in [27]. According to these measurements, we assume that the transmission rate of LTE and WiFi interfaces are uniformly and randomly distributed over $[4.85,\ 6.85]$ Mbps and $[2.01,\ 4.01]$ Mbps, respectively. We further assume that the transmission rate of Bluetooth is uniformly distributed in $[0.7,\ 2.1]$ Mbps. We also assume that the processing power of fog nodes is uniformly distributed in $[2,\ 3]$ GHz and the processing power of each cloud server is 4 GHz [29]. Unless stated otherwise, we assume that the roundtrip delay of remote cloud servers (i.e., $d^{\text{rt}}_{n,0}$) is 200 msec [8], while the roundtrip delay between IoT devices and fog nodes is negligible.

[2]For example, ARM Cortex–M3 processor which is widely used in IoT devices has 100 MHz clock speed [24]. Moreover, smartphones are usually powered with ARM Cortex–A8 processors with 1 GHz clock speed [25].

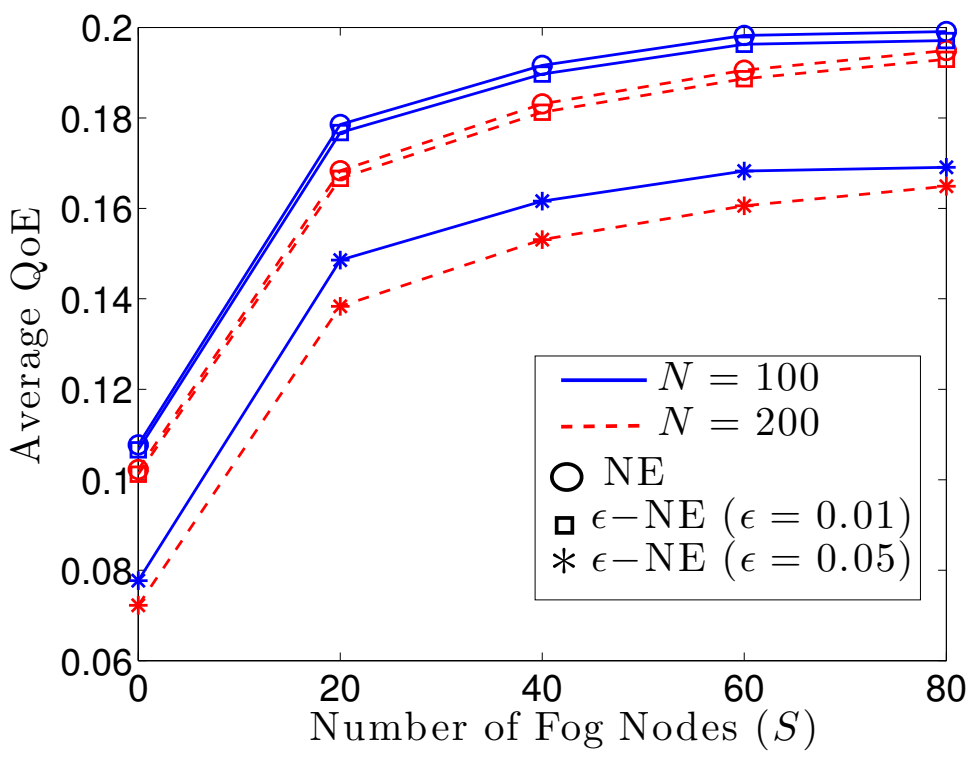

Fig. 2. The average perceived QoE of IoT users at the NE and $\epsilon$–Nash equilibria with different values of $\epsilon$.

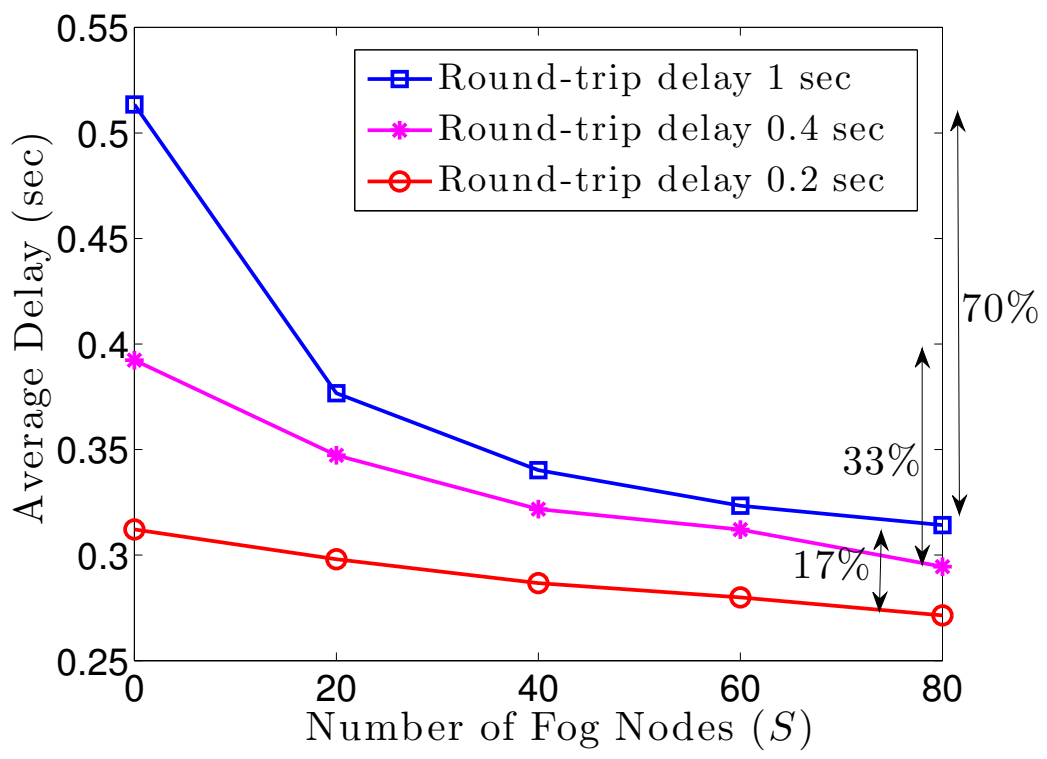

Fig. 3. The average delay each task experienced to be processed when there are $N = 200$ IoT users. We vary the roundtrip delay of remote cloud servers to study the computation time of delay–sensitive applications. When $S = 0$, the users can only offload their computation tasks to the cloud servers.

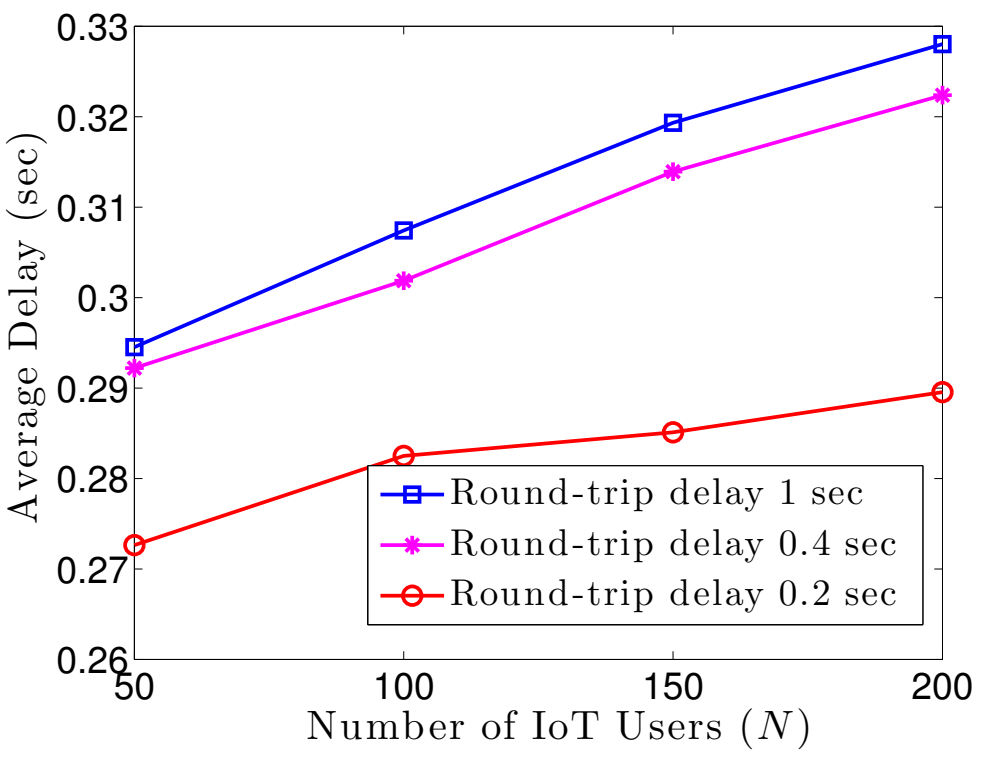

Fig. 4. The average delay each task experienced to be processed versus the number of IoT users $N$ when there are $S = 50$ fog nodes. The average delay significantly increases when there are more number of users. The users may not choose the fog nodes in this case due to their long computation time.

We consider different computing jobs with different task sizes and processing densities arrive at IoT users to study a wide range of IoT applications. We assume that the task size and processing density are uniformly distributed in $[100 \text{ B}, 0.5 \text{ MB}]$ and $[100, 600]$ cycles per bit, respectively. To study the tradeoff between energy consumption and delay, unless stated otherwise, we randomly choose $\lambda^{\text{E}}_n$ and $\lambda^{\text{T}}_n$ from the intervals $[0, 1]$ and $[0.5, 1]$, respectively.

## *B. Computation Offloading Game*

*1) Average QoE:* We first study the computation offloading game by investigating the average QoE of users at different equilibria obtained by Algorithm 1 (i.e., NE) and Algorithm 2 (i.e., $\epsilon$–Nash equilibria). The QoE of each user $n \in \mathcal{N}$ is given in (5). Fig. 2 illustrates the average QoE perceived by IoT users by utilizing the computing services. As can be observed, for $\epsilon = 0.01$, the QoE perceived at the $\epsilon$–Nash equilibrium is almost the same as what achieved at the NE. This is because when $\epsilon$ is very small, the users may choose the same strategy as the NE. However, as stated in Theorem 3, the convergence time of Algorithm 2 is guaranteed to be polynomial. By increasing the value of $\epsilon$, we trade the QoE for reducing the computational complexity.

Fig. 2 further shows that the average QoE of users in an IoT system with more fog nodes is higher than that of a system with fewer fog nodes. The more number of fog nodes, hence the more nearby computing resources, reduces the delay each offloaded task experiences, which consequently improves the QoE of users. However, when there are more IoT users, the workload of each fog node increases. Thus, the average QoE each user obtains reduces.

*2) Average Delay:* We further focus on delay–sensitive applications and investigate the delay each task experienced to either be processed locally or be offloaded. We set $\lambda^{\text{E}}_n = 0$ and $\lambda^{\text{T}}_n = 1$ for all $n \in \mathcal{N}$ and vary the roundtrip delay between IoT devices and the remote cloud servers. Fig. 3 shows the average delay when different number of fog nodes exist in the system. We first investigate the case that there are no fog nodes (i.e., $S = 0$) and the IoT users can only offload their computation tasks to the remote cloud servers. As can be observed, each task experienced a huge delay when $S = 0$. However, by increasing the number of fog nodes, the delay significantly reduces as the users can offload their computation tasks to their nearby fog nodes with low latency. This demonstrates the ability of fog nodes in providing low–latency computing

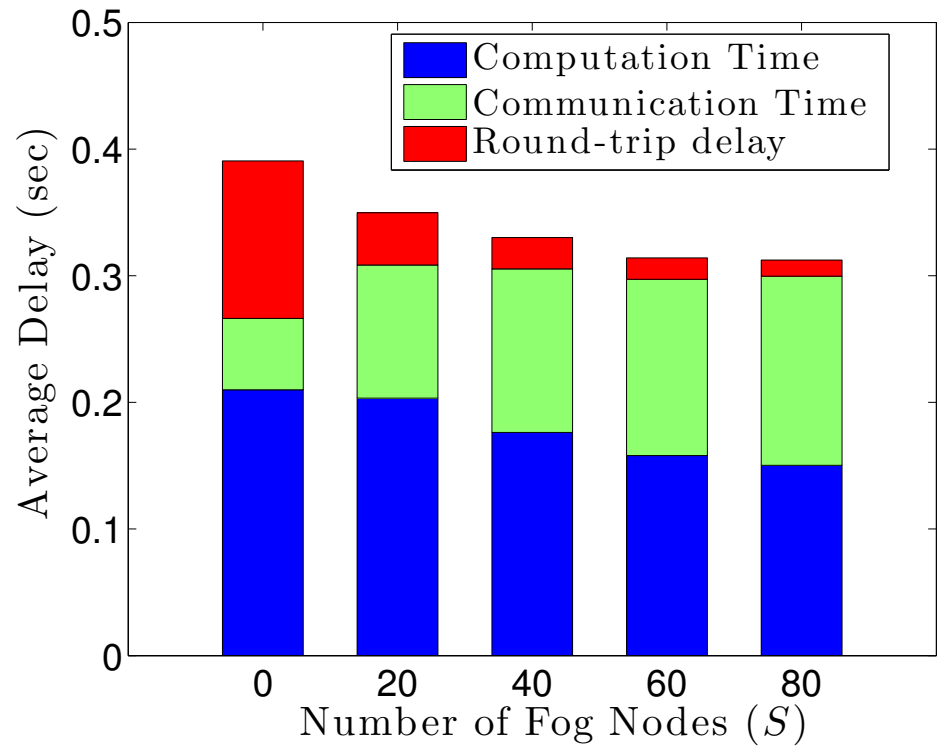

Fig. 5. The average delay each task experiences categorized into the computation time, communication time, and the roundtrip delay.

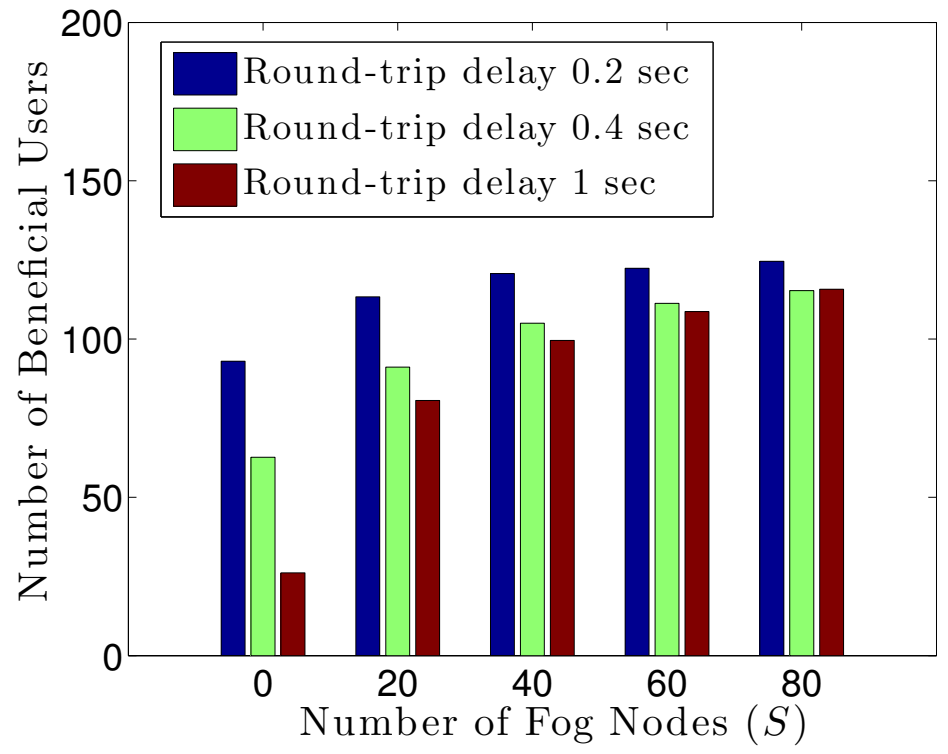

Fig. 6. The number of users that find the computing services beneficial and offload their tasks affected by the roundtrip delay. The QoE of these users are greater than zero ($N = 200$).

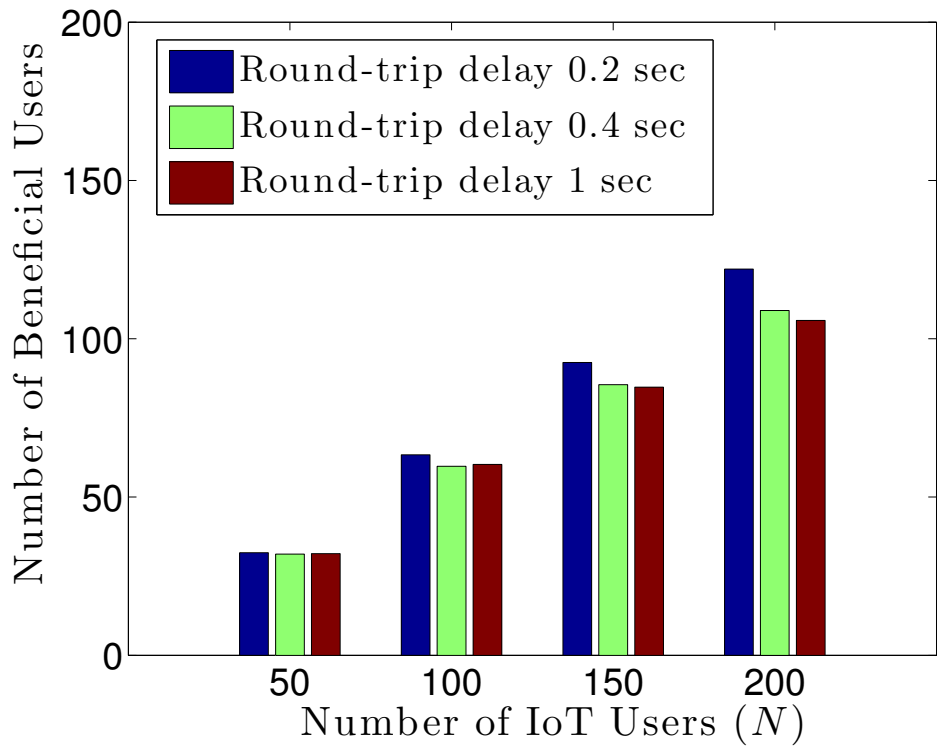

Fig. 7. The number of beneficial users affected by the roundtrip delay when $S = 50$.

services for delay–sensitive IoT applications.

Fig. 4 further shows the average delay of tasks when there are $S = 50$ fog nodes in the system. We vary the number of IoT users for different roundtrip delays of remote cloud servers. A larger roundtrip delay to offload the tasks to the remote cloud servers substantially increases the average delay each task experiences. This is because the users are not interested in the remote cloud servers anymore, while the processing power of fog nodes is also limited.

We also investigate the computation time, communication time, and the roundtrip delay that each task experiences. Fig. 5 shows the average of these metrics for different number of fog nodes when there are $N = 200$ IoT users in the system. As can be observed, the computation time reduces when there are more fog nodes. This is because more IoT users choose computation task offloading rather than local computing, which consequently reduces the computation time. Likewise, the roundtrip delay reduces in this case. However, the communication time increases as the WiFi and Bluetooth interfaces used for dispatching the tasks to the fog nodes have usually a lower transmission rate than the LTE interface used for transmitting the tasks to the remote cloud servers. Notice that in local computing, the communication time is zero as the tasks are performed locally.

*3) Number of Beneficial Users:* We now study the number of beneficial users that offload their computation tasks as their QoE is greater than zero. Figs. 6 and 7 show the number of beneficial users for different number of fog nodes and IoT users affected by the roundtrip delay of remote cloud servers. As can be observed from Fig. 6, when there are no fog nodes, the number of beneficial users substantially decreases for long roundtrip delay of remote cloud servers. However, the effect of roundtrip delay will be reduced when fog nodes are available for computation task offloading. According to Fig. 7, if there are more users in the system, a higher number of them benefit from the computing services. However, the percentage of beneficial users decreases as the workload of fog nodes and cloud servers, hence the computation time to perform the task, increases. For example, when there are 100 users and the roundtrip delay is 200 msec, 70% of users offload their tasks, while for a system with 200 users, only 58% of them have non-zero QoE. Surprisingly, when $N = 50$, the effect of different roundtrip delay is negligible. This is because there are enough nearby computing resources in this case and the IoT users do not offload their computation tasks to the remote cloud servers.

Figs. 8 and 9 illustrate the number of beneficial users categorized into the number of users offloading to the remote cloud servers and fog nodes, respectively. As shown in Fig. 8, the number of computation tasks offloaded to the remote cloud servers significantly reduces when there are more fog nodes offering the computing services. Furthermore, from Fig. 9, we can also observe that more users offload their computation tasks when we increase $N$. In this case, the increase in the number of tasks offloaded to the remote cloud servers is more significant comparing to that of fog nodes. This is because a higher workload at the fog nodes imposed by more number of offloaded tasks affects the interest of IoT users in utilizing fog computing services.

We now compare the performance of our proposed mechanism with that of [8][3] when the roundtrip delay is 400 msec. We assume that $\lambda_n^{\mathrm{E}} = 0$ for all $n \in \mathcal{N}$, while $\lambda_n^{\mathrm{T}}$ is randomly and uniformly distributed in $[0.01,\ 1]$ in order to compare in a fair manner. We consider $\lambda_n^{\mathrm{T}}$ for each task $n \in \mathcal{N}$ as its priority when investigating the performance of the offloading mechanism [8]. Figs. 10 and 11 illustrate the number of IoT users that benefit from computing services by utilizing our proposed computation offloading mechanism in comparison to [8]. As can be observed from Fig. 10, our proposed mechanism outperforms [8] by up to 20% in terms of the number of beneficial users. We further vary the number of fog nodes and investigate the performance of our proposed mechanism. As shown in Fig. 11, 18% more users benefit from computing services in comparision to the offloading mechanism proposed in [8]. Furthermore, Fig. 11 shows that the number of beneficial users significantly decreases when there is no fog node (i.e., $S = 0$) in the system.

[3]In [8], the authors proposed offline and online mechanisms. Since the offline mechanism always outperforms the online one, we used the offline mechanism for comparison. Notice that the offline mechanism in [8] aims to minimize the total delay of all tasks by considering their priorities.

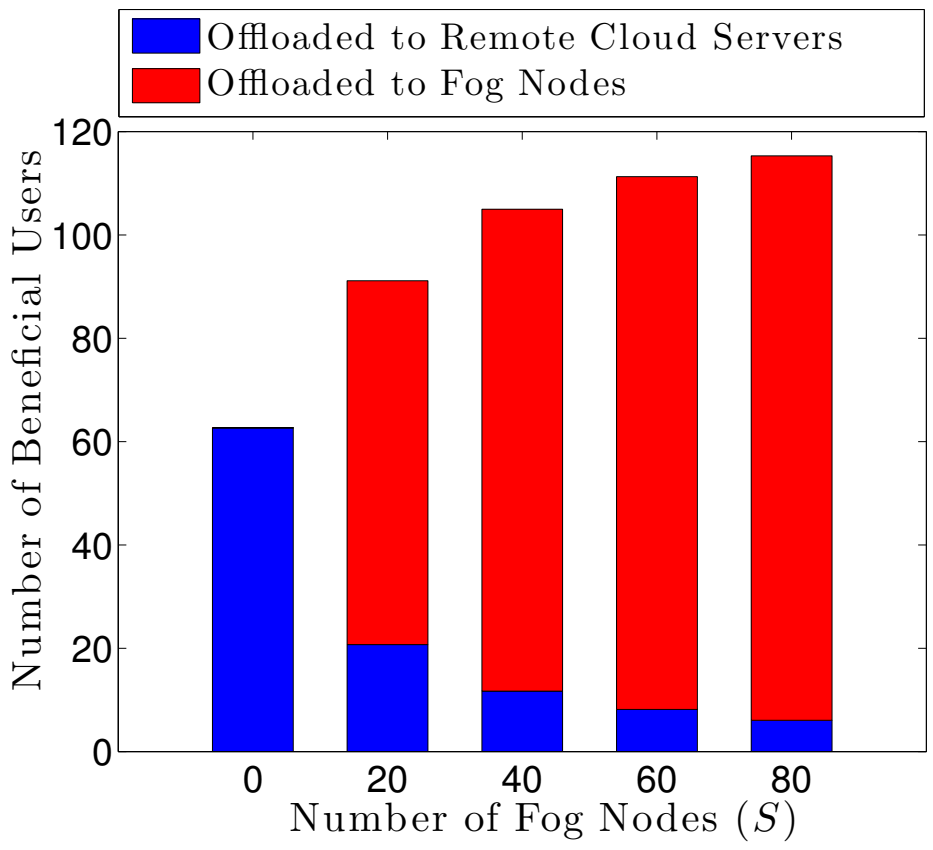

Fig. 8. The number of beneficial users categorized into the number of users offloading to the remote cloud servers and fog nodes, respectively ($N = 200$).

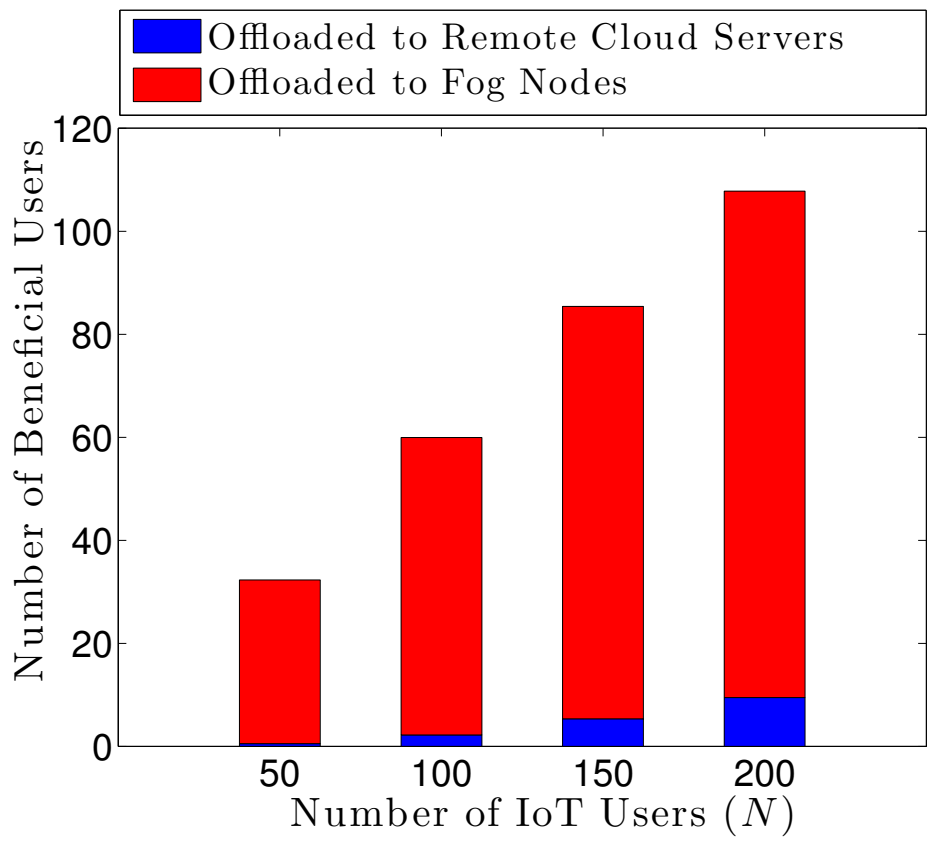

Fig. 9. The number of beneficial users categorized into the number of users offloading to the remote cloud servers and fog nodes, respectively ($S = 50$).

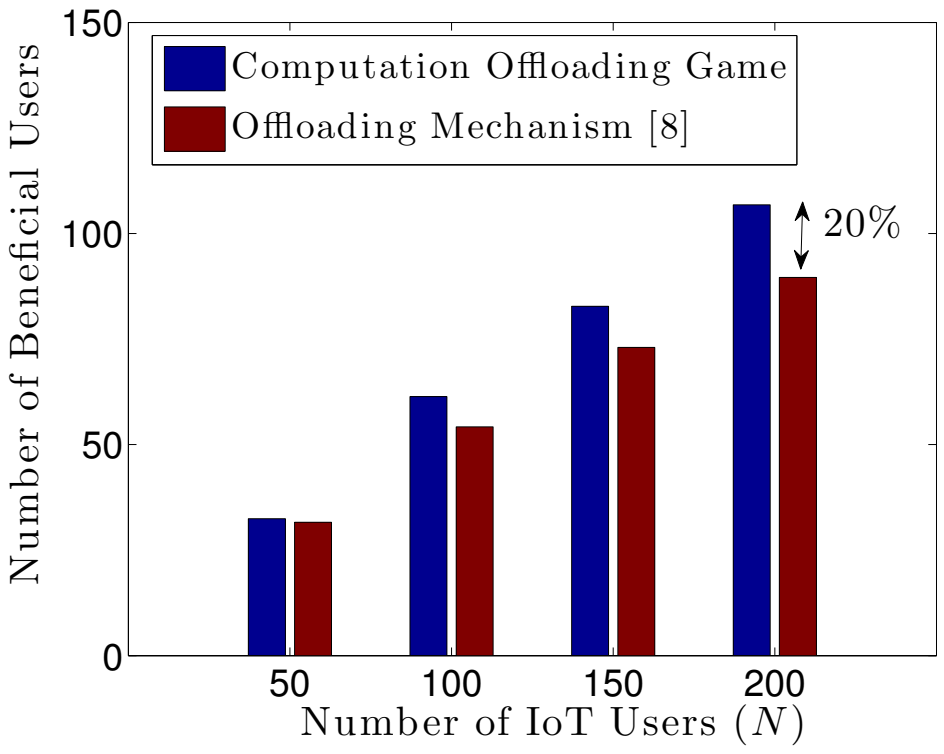

Fig. 10. The number of beneficial users versus the number of IoT users. Our proposed mechanism outperforms the offloading mechanism propoposed in [8] by up to 20 % in terms of the number of beneficial users ($S = 60$).

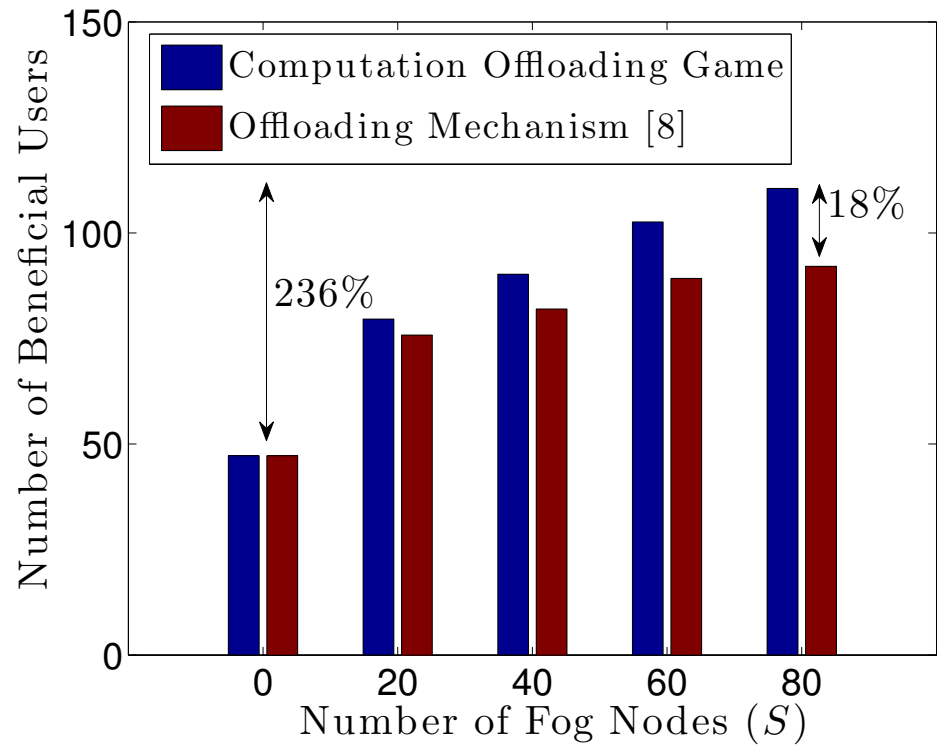

Fig. 11. The number of beneficial users versus the number of fog nodes ($N = 200$). When $S = 0$, only the remote cloud services are available.

### C. Social Cost and PoA

In this subsection, we study the social cost of IoT users to investigate the efficiency of the NE in Game $\mathcal{G}$. The social cost is the total cost of all users as given in (11). Fig. 12 shows the total cost imposed to IoT users in different scenarios. The local computing refers to the case when all tasks are performed locally by users, while in the remote cloud computing case, we assume that only remote cloud servers are available. The cost of users in the computation offloading game is $\Phi\left(\mathbf{A}^{\mathrm{NE}}\right)$ and the socially optimal cost is $\Phi\left(\mathbf{A}^{\mathrm{S}}\right)$. The socially optimal cost is always less than the cost in the computation offloading game due to the strategic behavior of IoT users. However, as can be observed, the total cost in the computation offloading game is close to optimal social cost. This shows that the PoA of Game $\mathcal{G}$ is close to 1, which states that the degradation of the social cost due to the strategic behavior of players is negligible. This further validates Theorem 2 stating that the PoA is no worse than a constant. Fig. 12 also illustrates that if all users perform their computation tasks locally, a huge amount of cost will be imposed to them. Thus, the proposed hierarchical fog-cloud computing paradigm can significantly reduce the total cost of the IoT users and is promising for the development of future IoT applications.

## VI. Conclusion

In this paper, we studied the computing resource allocation in a hierarchical fog-cloud computing paradigm. We formulated a QoE maximization problem based on which we proposed a computation offloading game to model the competition between IoT users. We proved the existence of a pure NE and developed an algorithm that can achieve the equilibrium. We also provided an upper bound on the equilibrium efficiency loss of the game. To mitigate the time complexity of obtaining the NE, we further proposed a near–optimal resource allocation algorithm and showed that it reaches an $\epsilon$-Nash equilibrium in polynomial time. We investigated the proposed algorithm through numerical experiments. Our results show that by utilizing the proposed algorithms, the IoT users can obtain a higher QoE. Results also show that the computation time of delay-sensitive IoT applications reduces significantly when utilizing the computing resources of fog nodes. This demonstrates the ability of fog nodes in providing low–latency computing services in IoT systems. We further showed that the number of users that find the computing services beneficial increases by up to 20% when using the proposed mechanism in comparison to an existing algorithm in the literature.

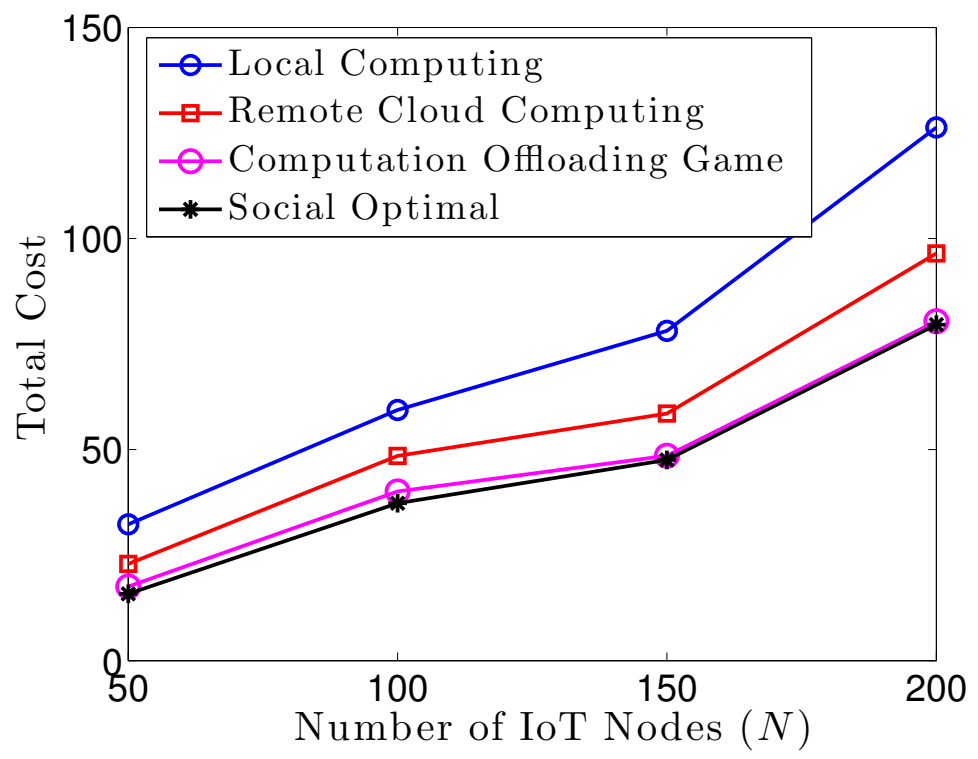

Fig. 12. The total cost of all IoT users. Our proposed computation offloading mechanism is able to achieve a close-to-optimal social cost.

For future work, we will consider dynamic arrival of computation tasks and develop an online mechanism to allocate the computing resources. In addition, we will focus on mobile edge computing, where the cloud servers are installed within the radio access networks of cellular systems. We will jointly allocate the wireless spectrum and computing resources to enable mobile users to offload their computation tasks. This is promising to realize low–latency edge network services of future 5G wireless networks.